# Skyrmion Brownian circuit implemented in a continuous ferromagnetic thin film

Yuma Jibiki,[†,‡] Minori Goto,[†,‡] Eiiti Tamura,[†,‡] Jaehun Cho,[†,‡,※] Soma, Miki, [†,‡] Hikaru Nomura,[†,‡] Titiksha Srivastava,[§] Willy Lim,[§] Stephane Auffret,[§] Claire Baraduc,[§] Helene Bea,[§] and Yoshishige Suzuki,[†,‡]*

[†] *Graduate School of Engineering Science, Osaka University, 1-3, Machikaneyamacho, Toyonaka, Osaka 560-8531, Japan*

[‡] *Center for Spintronics Research Network (CSRN), Graduate School of Engineering Science, Osaka University, 1-3, Machikaneyamacho, Toyonaka, Osaka 560-8531, Japan*

[§] *Univ. Grenoble Alpes, CEA, CNRS, Spintec, 38000 Grenoble, France*

[※] *Present Addresses: Korea Research Institute of Standards and Science (KRISS), 267 Gajeong-ro, Yuseong-gu, Daejeon 34113, Korea*

* Corresponding author: suzuki-y@mp.es.osaka-u.ac.jp

Abstract

The fabrication of a skyrmion circuit which stabilizes skyrmions is important to realize micro- to nano-sized skyrmion devices. One example of promising skyrmion-based device is Brownian computers, which have been theoretically proposed, but not realized. It would require a skyrmion circuit in which the skyrmion is stabilized and easily movable. However, the usual skyrmion circuits fabricated by etching of the ferromagnetic film decrease the demagnetization field stabilizing the skyrmions, and thus prevent their formation. In this study, a skyrmion Brownian circuit implemented in a continuous ferromagnetic film with patterned $SiO_2$ capping to stabilize the skyrmion formation. The patterned $SiO_2$ capping controls the saturation field of the ferromagnetic layer and forms a wire-shaped skyrmion potential well, which stabilizes skyrmion formation in the circuit. Moreover, we implement a hub (Y-junction) circuit without pinning sites at the junction by patterned $SiO_2$ capping. This technique enables the efficient control of skyrmion-based memory and logic devices, as well as Brownian computers.

Magnetic skyrmion is a topologically protected spin texture[1], which shows potential for implementation in the next generation of racetrack memory[2], logic[3], and neuromorphic devices[4]. The skyrmions were first observed in bulk MnSi at a low temperature[5], after which they were also observed at room temperature in FeGe thin film[6], Ta | CoFeB | TaO$_x$[7], and Ta | CoFeB | MgO[8] multilayers, the latter being conventionally used in magnetic tunnel junctions. Additionally, skyrmions can be electrically controlled by spin transfer torque[9-10], spin orbit torque[7], and electric field[11-14]. Furthermore, skyrmion Brownian motion has been investigated both theoretically[15-18] and experimentally[14,19-21].

Brownian motion has been investigated for calculation, such as in Brownian computing[22] and probabilistic computing[18]. A Brownian computer performs calculations using a small amount of energy close to the thermodynamic limit and the random motion of a Brownian particle for the carriage of information. The Brownian computer and its circuit architecture have been theoretically proposed[23-24], but not realized. Magnetic skyrmions are suitable for the Brownian computer because they act as Brownian particles in solid state materials and are electrically controllable and detectable at room temperature. Pinna *et al.*[18] and Zázvorka *et al.*[21] also proposed probabilistic computing using the stochastic properties of skyrmion Brownian motion. The realization of these applications requires a skyrmion circuit in which the skyrmion is stabilized and easily movable. In this study, we demonstrate a skyrmion Brownian circuit in continuous CoFeB film with patterned SiO$_2$ capping to stabilize the skyrmion formation. Moreover, we demonstrate a skyrmion hub (Y-junction) without pinning sites, which is a significant device used in Brownian computing[23-24].

The samples, Ta(5) | CoFeB(1.3) | Ta(0.22) | MgO(1.5) | SiO$_2$ (described in nm)[13,21], were deposited on thermally oxidized Si substrates by a magnetron sputtering system (see the Methods section for details). Figure 1(a) shows the perpendicular magnetic field dependence of the polar magneto-optical Kerr effect (MOKE) signal of continuous film. The colors represent the thickness of SiO$_2$ capping. We found that the saturation field decreased with the increase of SiO$_2$ thickness even though the SiO$_2$ was not in direct contact with the CoFeB layer. This result suggests that, through the strain, the magnetic anisotropy is affected by the SiO$_2$ deposition. Figures 1(b) and 1(c) show the MOKE microscope images of the sample with an SiO$_2$ thickness of 3.0 nm at the perpendicular magnetic field $H$ of 0 Oe and 2.8 Oe, respectively. The maze domain and skyrmions are observed at $H = 0$ and $H = 2.8$ Oe, respectively. At $H = 2.8$ Oe, the skyrmion Brownian motion was observed. All measurements were performed at a temperature of 303 K.

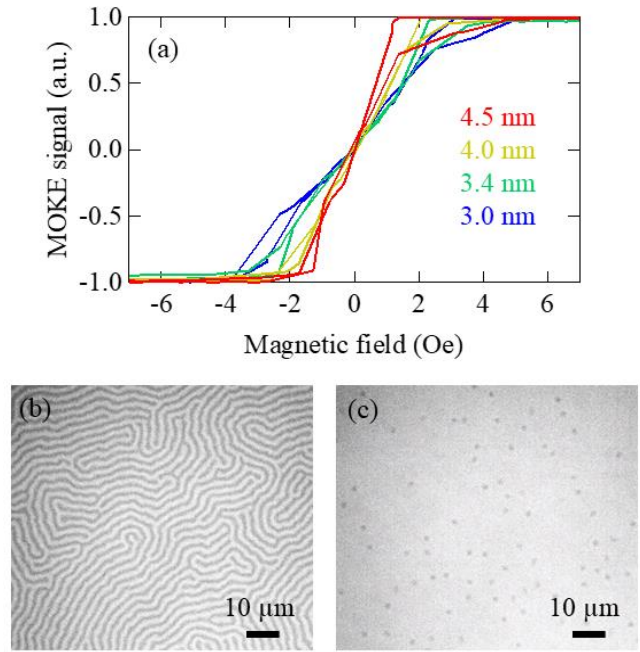

**Figure 1.** (a) Perpendicular magnetic field dependence of polar magneto-optical Kerr effect (MOKE) signal of the continuous film before process. Color shows the $SiO_2$ capping thickness. (b) (c) polar-MOKE microscope images of the sample with $SiO_2$ thickness of 3.0 nm at a temperature of 303 K and perpendicular magnetic field of (b) $H = 0$ and (c) $H = 2.8$ Oe. The scale bar is 10 μm.

We fabricated a wire-shaped sample using Ar ion milling and electron beam lithography. The $SiO_2$ capping thickness was 3.0 nm. Figure 2 shows the phase diagram of the magnetic domain in etched wire with various wire widths and a perpendicular magnetic field. We found that the decrease in wire width prevented the formation of skyrmions. This result suggests that the etching caused the decrease of the demagnetization field. The demagnetization field due to the opposing magnetization outside of the skyrmion stabilizes the skyrmion formation[25]. Therefore, removal of the magnetic material outside the wire by etching suppresses the skyrmion formation.

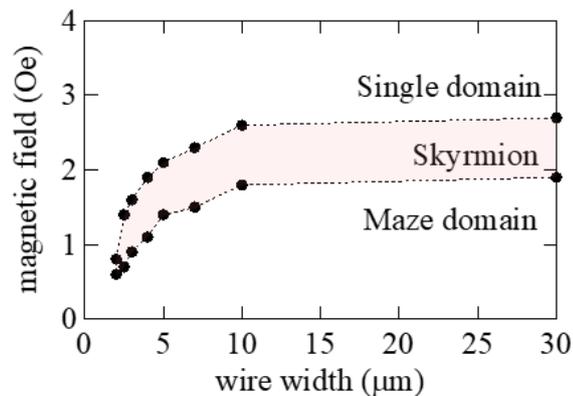

**Figure 2.** The phase diagram of magnetic domain in etched wires under various perpendicular magnetic fields and wire widths. The colored area represents the skyrmion phase.

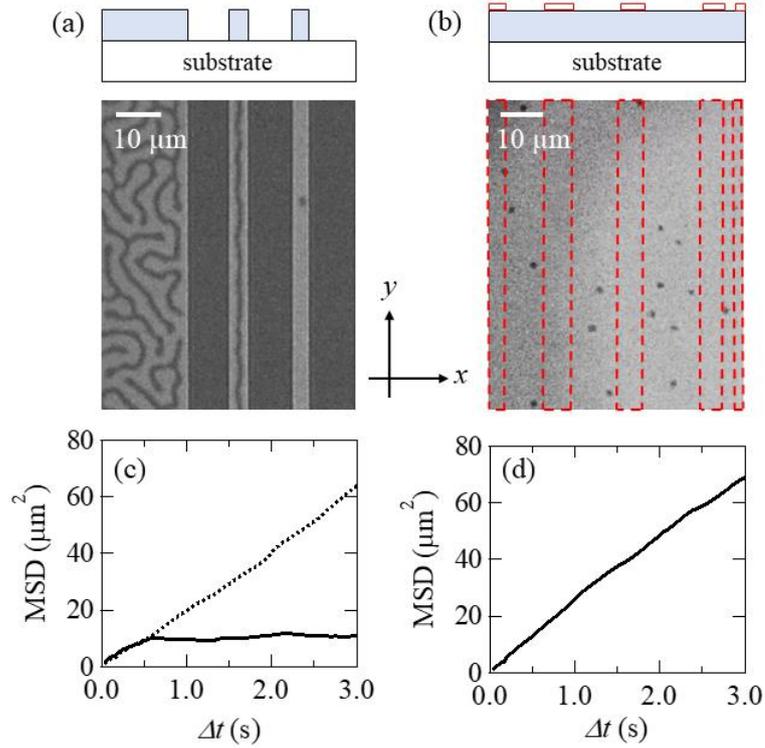

**Figure 3.** (a) MOKE microscope image and schematic of cross-sectional image of the etched film. Blue squares represent sputtered film including the CoFeB layer. The MOKE image is observed at the perpendicular magnetic field of 0.6 Oe and temperature of 303 K. (b) MOKE microscope image and schematic of cross-sectional image of the continuous film with patterned SiO$_2$ capping. Red solid and dashed squares show the estimated pattern of the sputtered SiO$_2$ with the thickness of 0.2 nm fabricated by electron beam lithography. The MOKE image is observed at the perpendicular magnetic field of 2.8 Oe and temperature of 303 K. (c) (d) Time difference dependence of mean square displacement (MSD) of skyrmion in the (c) etched wire and (d) wire formed by patterned SiO$_2$ capping on continuous film. The solid and dashed lines in (c) represent the pinned and unpinned skyrmion, respectively.

Following this, two types of samples, implemented by etching by Ar ion milling and depositing a patterned SiO$_2$ capping on continuous film, respectively, were compared (see the Methods section for details). One sample was etched in a wire shape down to the substrate layer by an electron beam lithography and the Ar ion milling. The other sample was produced by depositing SiO$_2$ with a thickness of 0.2 nm in the strip pattern on the 3.0 nm thick SiO$_2$ capping by electron beam lithography with the lift-off technique and magnetron sputtering system. Figures 3(a) and 3(b) show the schematics of the cross-sectional images (top) and the MOKE microscope images (bottom) of the etched film and the continuous film with patterned SiO$_2$ capping, respectively. In Figs. 3(a) and 3(b), the perpendicular magnetic fields of 0.6 Oe and 2.8 Oe, respectively, were applied. The extra 0.2 nm SiO$_2$ was deposited in the area surrounded by red dashed lines in shown Fig. 3(b). The skyrmion density is lower in this area because the saturation field is lower, *i.e.,* the magnetic potential energy of skyrmion is higher. As shown in Fig. 3(a), in etched samples, the magnetic domains strongly depend on the wire width from maze domain to skyrmion. In contrast, in the patterned SiO₂ capping samples, the skyrmions exist irrespective of the wire width. This shows that a continuous film with patterned SiO₂ capping stabilizes the skyrmion formation because of the presence of the demagnetization field. We characterize the dependence of the time difference $\Delta t$ on the one dimensional mean square displacement (MSD) in the etched film and continuous film with patterned SiO₂ capping, as shown in Figs. 3(c) and 3(d), respectively. The MSD is calculated as $\mathrm{MSD} = \langle [y(t+\Delta t) - y(t)]^2 \rangle = 2D\Delta t$. Here, $D$ is the diffusion constant, and $\Delta t$ is the time difference in the time series of skyrmion position $y$. MSD is averaged over the time in the range from $t_i$ to $t_f - \Delta t$, where $t_i$ and $t_f$ represent the initial and final time, respectively, of the time series. The total observation times $t_f - t_i$ in Figs. 3(c) and 3(d) are 30 s and 20 s, respectively. In Fig. 3(c), the solid and dashed lines, respectively, show the MSD when the skyrmion is pinned and not pinned. The MSD of the pinned skyrmion saturates at approximately 10 μm². From the linear fitting of the dashed and solid lines in Figs. 3(c) and 3(d), respectively, the diffusion constants in etched film and continuous film with patterned SiO₂ capping are evaluated as D = 10 μm²/s and D = 12 μm²/s, respectively. The obtained values were approximately 10 times higher than those previous obtained at 303 K[21]. The high diffusion coefficients may be attributed to the fact that our CoFeB films, that were not annealed, are not crystalized but rather amorphous. The amorphous CoFeB had no pinning sites due to grain boundaries, which could have enhanced the diffusion constant.

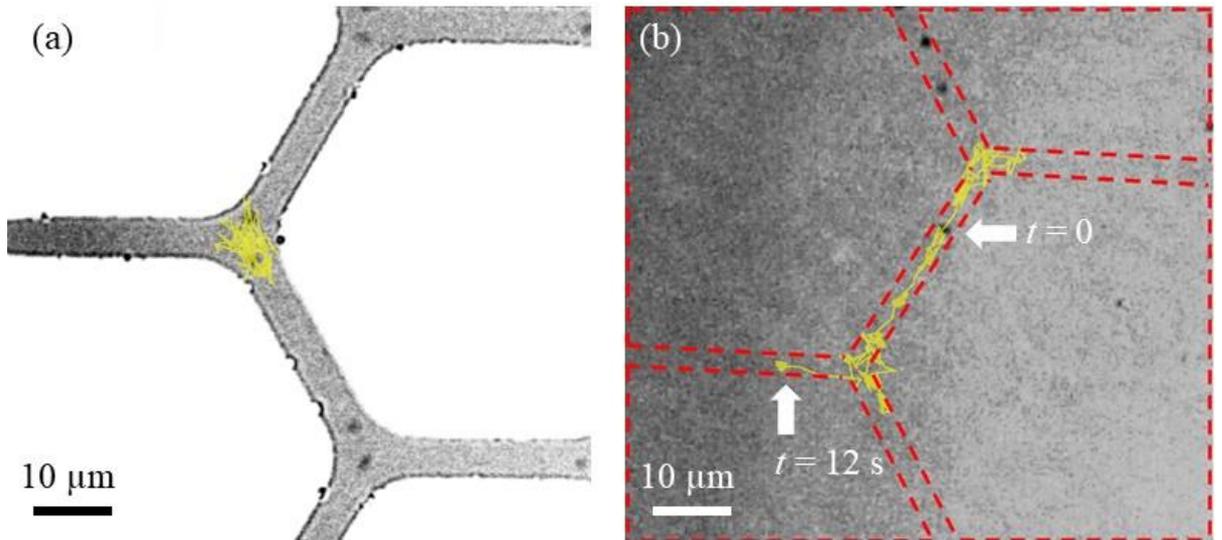

**Figure 4.** (a) (b) MOKE microscope images of the (a) etched hub and (b) hub formed by patterned SiO₂ capping on continuous film. The yellow lines depict the trajectory of the skyrmion. The SiO₂ with 0.2 nm thickness was deposited in the area enclosed by the red dashed line shown in (b) (estimated patterns). White arrows show the skyrmion position at the time $t = 0$ s and $t = 12$ s.

We demonstrate the skyrmion hub (Y-junction) by etched film and continuous film with patterned $SiO_2$ capping, as shown in Figs. 4(a) and 4(b), respectively. The yellow lines depict the trajectory of the skyrmion. The skyrmion in the etched hub is pinned at the center of the junction because the non-uniform demagnetization field forms a pinning potential. We propose that inside the hub, the distance to the edge is larger, more magnetic material with opposite magnetization surrounds the skyrmion, thus forming a pinning potential. In contrast, the skyrmion in the continuous film with patterned $SiO_2$ capping diffuses in the hub circuit without pinning. The results of this show that the continuous film with patterned $SiO_2$ capping efficiently removes the pinning sites.

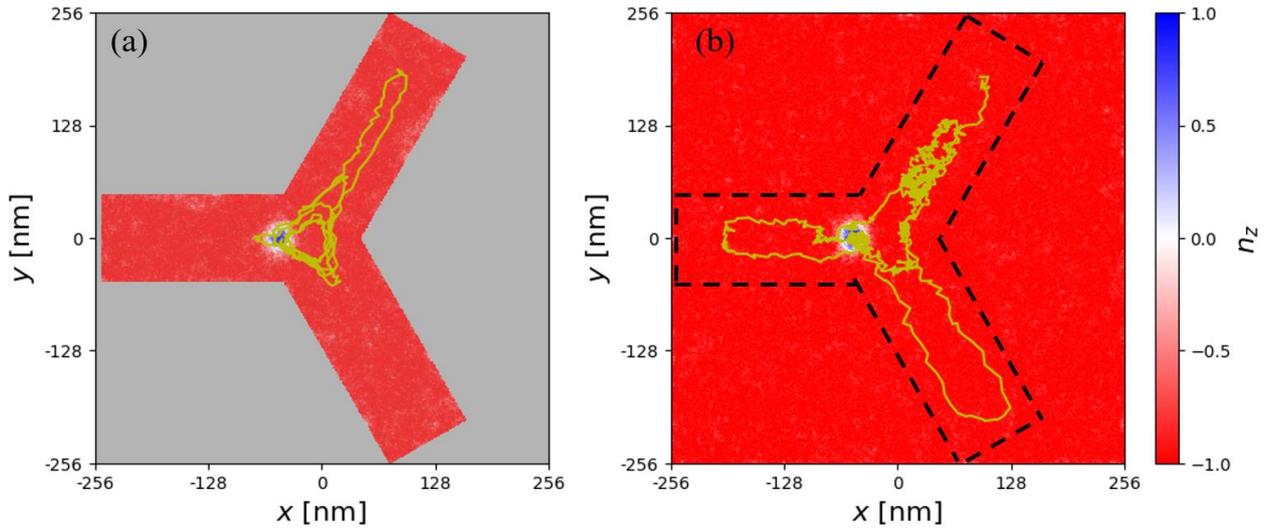

**Figure 5.** Micromagnetics simulation results of hub circuit in the etched hub (a) and the hub circuit patterned by magnetic anisotropy change (b), corresponding to the Figs. 4(a) and 4(b), respectively. $n_z$ is the normalized magnetization. Yellow lines represent the trajectories of the skyrmion position. Gray color in Fig. 5(a) represents no magnetization. The dashed line represents the boundary of anisotropy change.

This result is in good agreement with the micromagnetics simulation as shown in Fig. 5. Figures 5(a) and 5(b) show the simulation results of skyrmion hubs in the etched film and continuous film with patterned circuit by anisotropy change, respectively. We calculated the skyrmion dynamics by using a micromagnetics simulator MuMax3 with a GPU[1]. The simulation area is 512 nm × 512 nm × 1.2 nm, and the width of the hubs is 100 nm. The damping constant $α$ is 0.03, the saturation magnetization $M_s$ is 580 kA/m$^3$. The exchange stiffness constant $A$, Dzyaloshinskii-Moriya interaction (DMI) coefficient $D$, and perpendicular magnetic anisotropy $K$ in the circuit are 0.25 pJ/m, 1.9 mJ/m$^2$, and 0.4 MJ/m$^3$, respectively. The perpendicular magnetic anisotropy of the outside of the circuit in Fig. 5(b) is 0.6 MJ/m$^3$. Skyrmion is easily trapped at the junction of the etched hub as shown in Fig. 5(a), while that is not trapped in the hub circuit patterned by anisotropy change as shown in Fig. 5(b) because the inhomogeneous demagnetizing field is removed. Our result shows that the patterning circuit by anisotropy change enables the skyrmions to move each directions of the hub.

Finally, we measured the probability of right and left turns at the junction to observe whether or not skyrmion gyration affects the Brownian motion or not. The gyration of the skyrmion by current driven motion was confirmed as shown in Fig. 6; skyrmion has a transverse motion under the current. This result suggests that the current exerts the gyration force on the skyrmion. However, the probabilities of right and left turns at the hub junction are both 50±17%. In this study, no significant difference in these probabilities was observed.

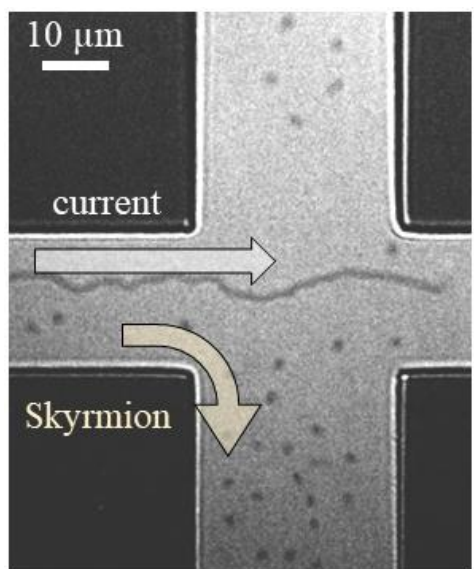

**Figure 6.** MOKE microscope image of the current driven skyrmion. Skyrmions move along transverse to the current direction.

In this study, a skyrmion Brownian circuit implemented in the continuous film with patterned $SiO_2$ capping was demonstrated. It was found that the skyrmion wire and hub circuit can be implemented using patterned $SiO_2$ capping, which controls the saturation field. Furthermore, in the hub circuits implemented by patterned $SiO_2$ capping on continuous film, the pinning site, usually observed in etched hubs, is removed. Thus, by this method, a hub (Y-junction) circuit can be implemented without pinning sites. The results of this work form a basis for skyrmion-based technology, especially in realizing Brownian computer.

METHOD

The samples, Ta(5) | CoFeB(1.3) | Ta(0.22) | MgO(1.5) | $SiO_2$ (described in nm), were deposited on thermally oxidized Si substrates by a magnetron sputtering system (Canon

ANELVA, E-880S-M in Osaka University). The two types of samples were implemented by etching by Ar ion milling and depositing a patterned $SiO_2$ capping on continuous film, respectively. One sample was etched in a wire shape down to the substrate layer by an electron beam lithography and the Ar ion milling. The other sample was fabricated by starting from the continuous film with 3 nm thick $SiO_2$ capping, then patterning strips in resist by electron beam lithography and depositing 0.2 nm $SiO_2$ film with magnetron sputtering, that was finally removed outside the strips by lift off technique. The resist on the sample was baked at 130 ℃ for microfabrication, which does not affect magnetic properties.

Skyrmions were observed by polar MOKE system. The magnetic field was applied by a permanent magnet. Measurements were performed at a temperature of 303 K controlled by a heater.

ACKNOWLEDGEMENT

This research and development work was supported by the French ANR ELECSPIN (contract no. ELECSPIN ANR-16-CE24-0018) and the Ministry of Internal Affairs and Communications.